\documentstyle[12pt,epsbox]{article}
\newcommand\beq{\begin{eqnarray}}
\newcommand\eeq{\end{eqnarray}}

\renewcommand{\theequation}{\thesection.\arabic{equation}}
\newcommand{\do}{^{\circ }}

\newcommand{\cl}{\centerline}

\def\0vec{\mbox{\boldmath $0$}}

\begin{document}
\setlength{\baselineskip}{0.2in}

\Large

\centerline{\bf NLO $Q^2$-evolution of the nucleon's}
\centerline{\bf transversity distribution $h_1(x, Q^2)$
\footnote{In Proceedings of the Workshop ``Deep Inelastic Scattering
off Polarized Targets: Theory Meets Experiment'', DESY-Zeuthen, Germany.
Sep. 1-5, 1997. (Ed. J. Bl\"{u}mlein et al. DESY 97-200) pages 157-166.
(presented by Y. Koike)}
}

\normalsize

\vspace{0.5cm}

\centerline{A. Hayashigaki, Y. Kanazawa and Yuji Koike}
\centerline{\it  Graduate School of Science and Technology,
Niigata University}
\centerline{\it Ikarashi, Niigata 950-21, Japan}

\vspace{0.5cm}

\centerline{\bf Abstract}
We present a calculation of the two-loop anomalous dimension
for the transversity distribution $h_1(x,Q^2)$, 
$\gamma^{h(1)}_n$, in 
the MS scheme of the dimensional regularization.  
Because of the chiral-odd nature, $h_1$ does not mix with
the gluon distributions, and thus our result is
the same for the flavor-singlet and nonsinglet distributions.
At small $n$ (moment of $h_1$),
$\gamma^{h(1)}_n$ is 
significantly larger than $\gamma^{f,g(1)}_n$ (the
anomalous dimension for the nonsinglet $f_1$ and $g_1$), 
but approaches $\gamma^{f,g(1)}_n$
very quickly at large $n$, keeping the relation
$\gamma^{h(1)}_n > \gamma^{f,g(1)}_n$.
This feature is in parallel to the relation between
the one-loop anomalous dimension for $f_1$ $(g_1)$ and $h_1$.  
We also show that
this difference in the anomalous dimension 
between $h_1$ and $g_1$ leads to
a drastic difference in the $Q^2$-evolution of those distributions
in the small $x$ region.

\section{Introduction}
\setcounter{equation}{0}
\renewcommand{\theequation}{\arabic{section}.\arabic{equation}}
\hspace*{\parindent}

The transversity distribution $h_1(x,Q^2)$ is the third and the final twist-2
quark distribution function of the nucleon.
It measures the 
probability in a tranversely polarized nucleon to find
a quark polarized parallel to the nucleon spin minus the probability
to find it oppositely polarized.
It thus supplies an information about the nucleon spin distribution
not provided by the $g_1$ distribution.
Because of its chiral-odd nature, $h_1$ can not be measured in the 
totally inclusive deep inelastic scattering (DIS), but it appears
as a leading contribution in the tranversely polarized
nucleon-nucleon Drell-Yan process\,\cite{RS,AM,JJ,CPR}, 
and the semi-inclusive
DIS which detects pions\,\cite{JJ}, $\Lambda$'s\,\cite{AM,J96}, and 
correlated two pions\,\cite{JJT}
in the final states.

Since $h_1$ is twist-2, its $Q^2$ evolution 
is described by the DGLAP equation\,\cite{DGLAP}.  The leading
order (LO) splitting function 
for $h_1$ has been known for some time\,\cite{AM}.
In the recent literature, three papers\,\cite{Vog, HKK, KM} discussed the
next-to-leading order (NLO) $Q^2$ evolution of $h_1$:
Vogelsang\,\cite{Vog} carried out the
light-cone gauge calculation of the two-loop splitting function\,\cite{CFP}
for $h_1$ and converted it into the anomalous dimension.
The present authors presented the Feynman gauge calculation of
the two-loop anomalous dimension\,\cite{HKK}.  
The final results of these 
two calculations in different formalisms agreed.
These results were
subsequently confirmed by \cite{KM} in the same formalism as \cite{HKK}.
These studies completed the whole list of the NLO anomalous dimension
for the nucleon's all twist-2 distributions together with the known result
for $f_1$\,\cite{FRS,GLY,CFP} and $g_1$\,\cite{MN}.

In this contribution, we discuss the NLO $Q^2$-evolution 
of $h_1$ following our paper \cite{HKK}.  We also present
a numerical solution of the NLO evolution equation for $h_1$,
comparing it with that of the $g_1$ distribution.  This latter part is 
a new result not presented in \cite{HKK}.

\section{Twist-two operator for $h_{1}(x,Q^{2})$ and its 
renormalization}
\renewcommand{\theequation}{\arabic{section}.\arabic{equation}}
\setcounter{equation}{0}

The transversity distribution $h_1(x,Q^2)$ 
of the nucleon is defined by the following
relation\,\cite{JJ}:
\begin{eqnarray}
& &\int{ d\lambda \over 2\pi}e^{i\lambda x} \langle PS | \bar{\psi}(0)
\sigma_{\mu\nu}i\gamma_5 \psi(\lambda n) |_Q |PS\rangle
= 2 \left[ h_1(x,Q^2)(S_{\perp\mu}p_\nu - S_{\perp\nu}p_\mu)/M 
\right.\nonumber\\
& &\left.\qquad\qquad + h_L(x,Q^2)M(p_\mu n_\nu - p_\nu n_\mu )
(S\cdot n) + h_3(x,Q^2)M (S_{\perp\mu}n_\nu - S_{\perp\nu}n_\mu) 
\right],
\label{eq2.1}
\end{eqnarray}
where $|PS\rangle$ is the nucleon (mass $M$) state specified by 
the four momentum
$P$ and the spin vector $S$, and two light-like vectors
$p$ and $n$ are introduced by the relation,
$P^\mu= p^\mu + {M^2 \over 2}n^\mu$ and $pn=1$.
$S^\mu$ is decomposed as $S^\mu= (S\cdot n)p^\mu + (S\cdot p)n^\mu
+ S_\perp^\mu$.  In (\ref{eq2.1}), the gauge link operator
to ensure the gauge invariance is suppressed.
$h_L$ and $h_3$ are twist-3 and -4 distributions which are not
in our interest here.  The quark distibutions $h_{1,L,3}$ in (\ref{eq2.1})
are defined for each quark flavor. 
The replacement of $\psi \to C\bar{\psi}^T$ and $\bar{\psi}
\to -\psi^T C^{-1}$ defines anti-quark distributions $\bar{h}_1(x,Q^2)$, 
and it is related to $h_1(x,Q^2)$ by the relation
$\bar{h}_1(x,Q^2) = -h_1(-x,Q^2)$.
Taylor expanding the both sides of (\ref{eq2.1}) with respect to 
$\lambda$,
one can relates the $n$-th moments of $h_{1,L}$ to
the following local operator:
\begin{eqnarray}
\theta_{\mu\nu\mu_{1}\ldots\mu_{n}} = {\cal S}_{n}\overline{\psi}
i\gamma_{5}\sigma_{\mu\nu}iD_{\mu_{1}}\ldots iD_{\mu_{n}}\psi,
\label{eq2.2}
\end{eqnarray}
where ${\cal S}_n$ denotes the symmetrization among $\nu,
\mu_1,\cdots,\mu_n$,
and the covariant derivative $D_\mu = \partial_\mu -igT^a A^a_\mu$
restores the gauge invariance.
In particular, the $n$-th moment of $h_1$
is related to a tower of the twist-2 operators:
\begin{eqnarray}
& &{\cal M}_n[h_1(Q^2)]\equiv\int_{-1}^1\,dx\,x^nh_1(x,Q^2)
=\int_{0}^1\,dx\,x^n \left[ h_1(x,Q^2) + (-1)^{n+1} \bar{h}_1(x,Q^2)
\right],
\label{eq2.3}\\[5pt]
& &\langle PS | 
\overline{\theta}_{\mu\nu\mu_1\cdots\mu_n}(Q)\Delta^\nu
\Delta^{\mu_1}\cdots \Delta^{\mu_n}
| PS \rangle
\nonumber\\ 
& &\qquad\qquad
=\frac{2}{M} 
{\cal M}_n[h_1(Q^2)]
\left(S_{\mu}\hat{P}^{n+1}-P_{\mu}
\hat{S}\hat{P}^{n}
      +\frac{2}{n+2}M^{2}\Delta_{\mu}
         \hat{S}\hat{P}^{n-1}\right),
\label{eq2.4}
\end{eqnarray}
where $\overline{\theta}_{\mu\nu\mu_1\cdots\mu_n}$ is defined 
as the traceless part of
$\theta_{\mu\nu\mu_1\cdots\mu_n}$ defined by the condition,
\begin{eqnarray}
g^{\mu\mu_{i}}\overline{\theta}_{\mu\nu\mu_{1}\ldots\mu_{n}}
=g^{\nu\mu_{i}}\overline{\theta}_{\mu\nu\mu_{1}\ldots\mu_{n}}
=g^{\mu_{i}\mu_{j}}\overline{\theta}_{\mu\nu\mu_{1}\ldots\mu_{n}}=0,
\label{eq2.5}
\end{eqnarray}
and we introduced a null vector $\Delta^\mu$ ($\Delta^2=0$) 
which symmetrizes the indices $\nu,\mu_1,\cdots,\mu_n$ and
kills off most of the trace terms in 
$\overline{\theta}_{\mu\nu\mu_1\cdots\mu_n}$ as usual.

For simplicity we symbolically use the notation $O_n$ for
$\overline{\theta}_{\mu\nu\mu_{1}\cdots\mu_{n}}$ in what follows
in this section.
The bare-($O_n^B$) and the renormalized-
($O_n(\mu)$) composite operators are related by the renormalization 
constant $Z_{n}(\mu)$ for $O_n$ as
\begin{eqnarray}
O_{n}(\mu)&=&Z_n^{-1}(\mu)O_{n}^{B}.
\label{eq2.6}
\end{eqnarray}
The scale dependence of $O_n(\mu)$ is obtained by solving the
renormalization group equation
\begin{eqnarray}
\mu\frac{dO_n(\mu)}{d\mu} + \gamma_n(g(\mu))
O_n(\mu)= 0,
\label{eq2.7}
\end{eqnarray}
where $\gamma_n(g(\mu))$ is the anomalous dimension
for the operator $O_n(\mu)$ defined as
\begin{eqnarray}
\gamma_n(g(\mu)) = \mu\frac{\partial}{\partial \mu}{\rm ln}
Z_n(\mu). 
\label{eq2.8}
\end{eqnarray}
This equation is solved to give
\begin{eqnarray}
O_n(Q^{2}) = O_n(\mu^{2}){\rm exp}
\left[-\int_{g(\mu^{2})}
^{g(Q^{2})}dg
\frac{\gamma_n(g)}{\beta(g)}\right]. 
\label{eq2.9}
\end{eqnarray}
Up to next-to-leading order, the anomalous dimension
$\gamma_n(\mu)$ and the beta function $\beta(g)$ can be expanded as
\begin{eqnarray}
\gamma_n(g)&=&{g^{2}\over 16\pi^2}\gamma_n^{(0)}
+{g^{4}\over (16\pi^2)^2}\gamma_n^{(1)}+O(g^{6}), 
\label{eq2.10}\\
\beta(g)&=&- {g^{3}\over 16\pi^2} \beta_0
-{g^{5}\over (16\pi^2)^2} \beta_1 
+O(g^{7}),
\label{eq2.11}
\end{eqnarray}
where the coefficients of the $\beta$-functions
are well known;  $\beta_0=11- (2/3)N_f$, $\beta_1=102-(38/3)N_f$,
with the number of quark flavor $N_f$.
Inserting these equations into (\ref{eq2.9}), one obtains
the next-to-leading order evolution equation for the $n$-th moment
of $h_1$ as
\begin{eqnarray}
{\cal M}_n[h_{1}(Q^{2})]&=&R_n(Q^2,\mu^2){\cal M}_n[h_{1}(\mu^{2})],
\label{eq2.12a}\\[5pt]
R_n(Q^2,\mu^2)&=&
\left({\alpha_s(Q^2) 
\over \alpha_s(\mu^2)}\right)^{\gamma^{(0)}_n/2\beta_0}
\left[ 1 +
{\alpha_s(Q^2) - \alpha_s(\mu^2) \over 4\pi}
{\beta_1\over \beta_0}
\left(\frac{\gamma_{n}^{(1)}}
{2\beta_{1}}-\frac{\gamma_{n}^{(0)}}{2\beta_{0}}\right)
\right],
\label{eq2.12}\\[5pt]
{\alpha_s(Q^2)\over 4\pi}&=&{1 \over \beta_0{\rm ln}(Q^2/\Lambda^2)}
\left[ 1- { \beta_1 {\rm ln}{\rm ln}(Q^2/\Lambda^2)
\over \beta_0^2 {\rm ln}(Q^2/\Lambda^2) } \right].
\label{eq2.12b}
\end{eqnarray}
The LO evolution equation can be obtained by putting
$\beta_1$ and $\gamma^{(1)}_n$ equal to zero in (\ref{eq2.12}) 
and (\ref{eq2.12b}).
This NLO effect in $h_1$ has to be combined with
the NLO effect in the hard cross section in the parton level
to give a prediction for a relevant physical
cross section in the NLO level.

\section{Two-loop anomalous dimension for $h_1(x,Q^2)$}
\setcounter{equation}{0}
\renewcommand{\theequation}{\arabic{section}.\arabic{equation}}

To obtain the renormalization constants for 
$\bar{\theta}^{\mu\nu\mu_1\cdots
\mu_n}$ in (\ref{eq2.4}), we calculate the two-loop correction
to the two-point function for $\bar{\theta}$
in the Feynmann gauge 
with off-shell quark lines having the momentum
$p$ ($p^2 < 0$).  Those diagrams are shown in Fig. 1.
Owing to the chiral-odd nature of 
$h_1$, it does not mix with the gluon distribution.
Therefore Fermion loop diagrams peculiar to
the flavor-singlet distribution do not contribute to  
the renormalization of $h_1$.
The calculation of the two-loop Feynman diagrams
is rather involved, and we refer 
the details of the calculation to our paper \cite{HKK}.

\newpage
\begin{figure}[h]
%\begin{center}
\epsfile{file=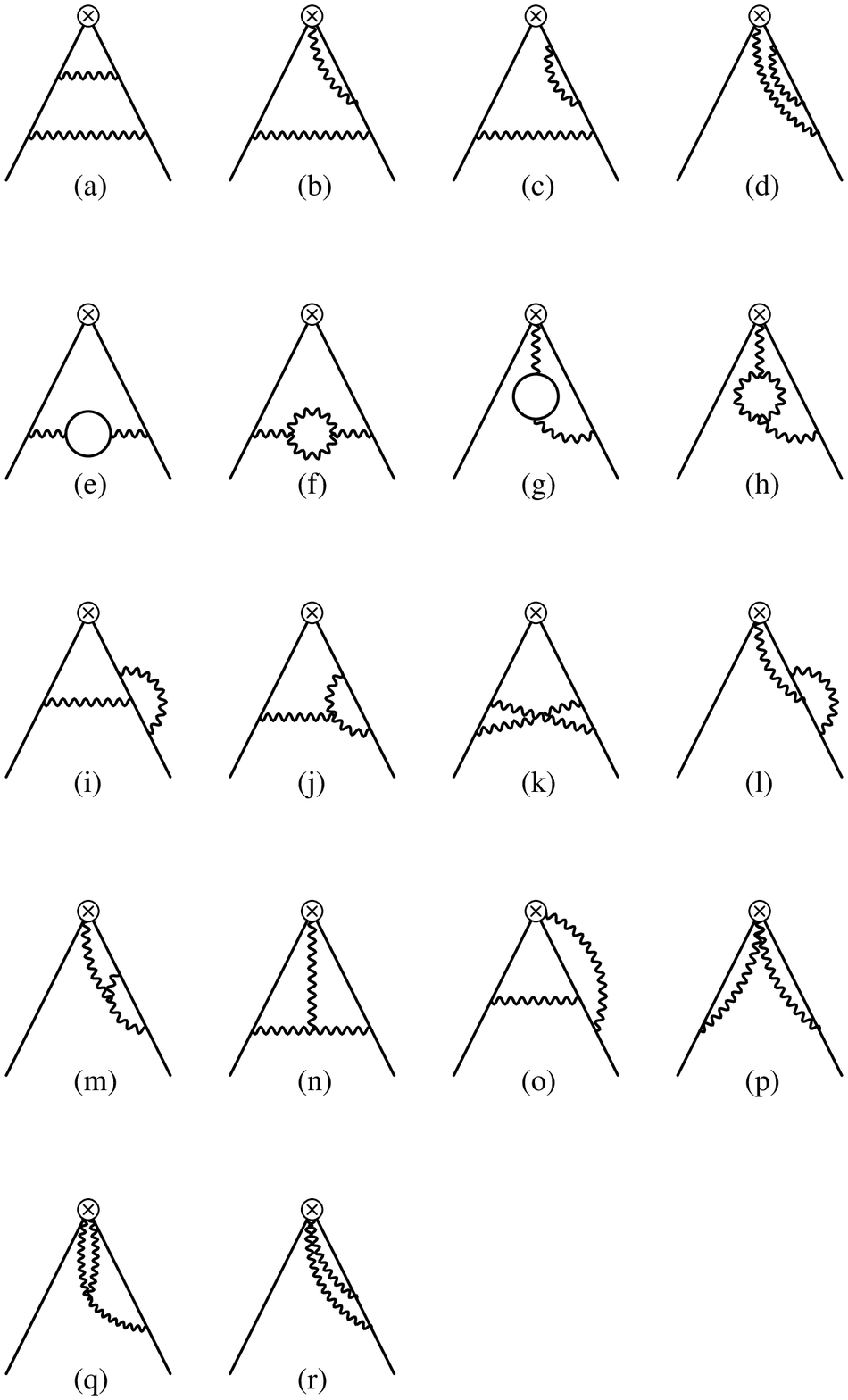,scale=0.4}
%\end{center}
%\caption[]{}
%\protect{\cite{}}}
%\label{figure:fignloA}
\end{figure}

\noindent
Fig. 1 : Two-loop corrections to
the one-particle irreducible
two-point Green function which imbeds the operator
$O_n^\mu\cdot\Delta$.
%For diagrams not symmetric, crossed diagrams must be included.
%In Figs. (f) and (h), the contributions from the Faddev-Popov ghost and
%tadpole diagram must be added. 

\vspace*{0.2in}

In the MS scheme, the final result for the two-loop anomalous dimension 
$\gamma^{(1)}_n$ (\ref{eq2.10}) for $h_1$ reads 
\begin{eqnarray}
\gamma^{h(1)}_n &=& 
4C_F^2 \left[ S_2(n+1)-2S_1(n+1) -{1\over 4}\right]\nonumber\\[5pt]
& &+C_FC_G\left[-16S_1(n+1)S_2(n+1) -
{58\over 3}S_2(n+1) + {572\over 9}S_1(n+1) -{20\over3}\right]
\nonumber\\[5pt]
& &-8\left(C_F^2 -{1\over 2}C_FC_G\right)
\left[4S_1(n+1)\left\{S_2'\left({n+1\over 2}\right)
-S_2(n+1) -{1\over 4}\right\}
-8\widetilde{S}(n+1)\right.\nonumber\\[5pt]
& &\left. + S_3'\left({n+1\over 2}\right)
-{5\over 2}S_2(n+1) + {(1+(-1)^n)\over (n+1)(n+2)} +{1\over 4}
\right]\nonumber\\[5pt]
& &
+{32\over 9}C_FT_R\left[3S_2(n+1)-5S_1(n+1) + {3\over 8}\right],
\label{eq4.0}
\end{eqnarray}
where
\begin{eqnarray}
S_k(n)&=&\sum_{j=1}^n{1\over j^k},\\[5pt]
S_k'\left({n\over 2}\right)&=&{1+(-1)^n \over 2}S_k\left({n\over 2}\right)     
+{ 1-(-1)^n \over 2}S_k\left({n-1\over 2}\right),\\[5pt]
\widetilde{S}(n)&=&\sum_{j=1}^n{ (-1)^j \over j^2}S_1(j),
\end{eqnarray}
and $C_F={N_c^2-1 \over 2N_c}$, $C_G=N_c$, $T_R = N_f/2$ with the 
number of color $N_c$ and the number of quark-flavor $N_f$.
Equation (\ref{eq4.0}) shows
$\gamma^{h(1)}_n \sim {\rm ln}n$ at large $n$ as was the case for 
$\gamma^{f(1)}_n$ ($\gamma^{(1)}_n$ for the nonsiglet $f_1$
and $g_1$)\,\cite{GLY}.
To compare their behavior, 
we plotted $\gamma^{h,f(1)}_n$ for $N_f=3,\ 5$ in Fig. 2.

\begin{figure}[h]
%\begin{center}
\epsfile{file=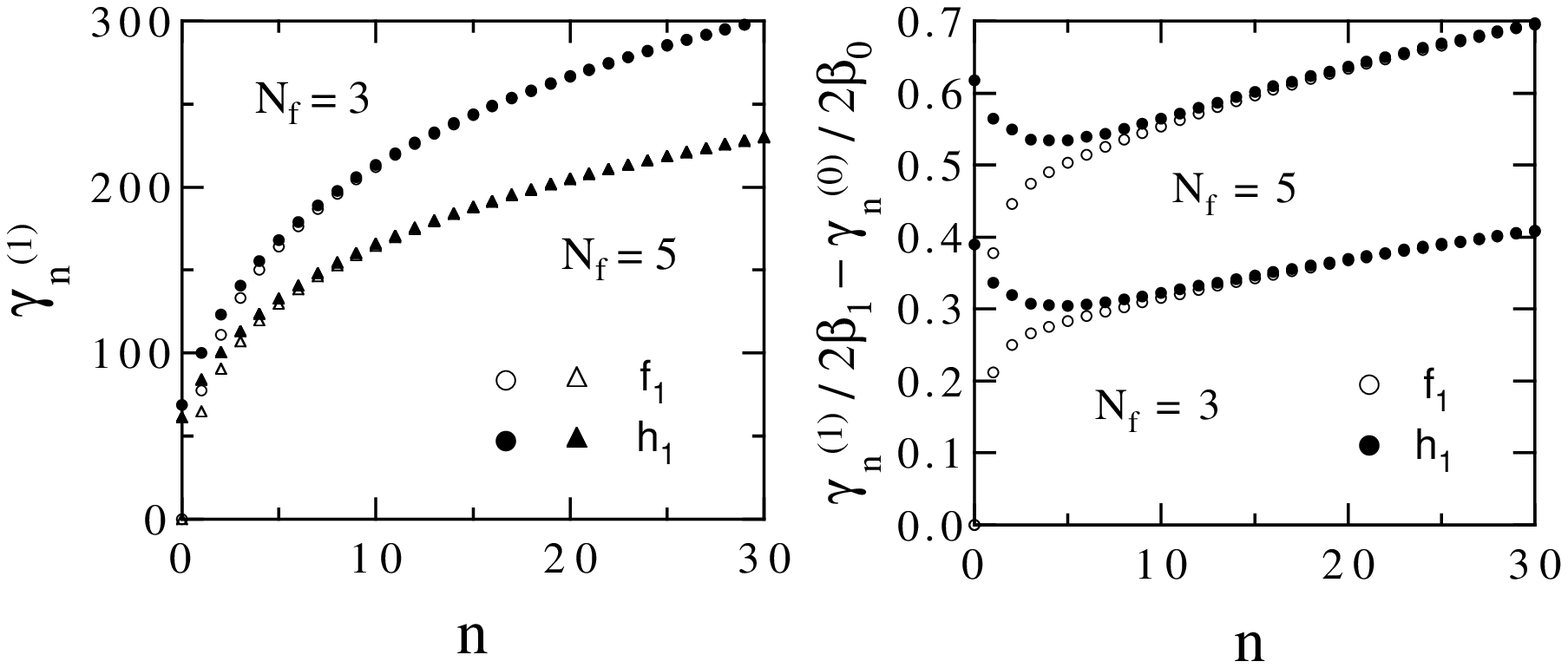,scale=0.8}
%\end{center}
%\caption[]{}
%\protect{\cite{}}}
%\label{figure:fignloB}
\end{figure}
%\hspace*{0.0in} 

\noindent
\begin{tabular}{lcccl}
Fig. 2 : Two-loop 
anomalous dimension & & &   &
Fig. 3 : $\gamma^{f,h(1)}_n/2\beta_1 
- \gamma^{f,h(0)}_n/2\beta_0$ for \\
$\gamma^{f,h(1)}_n$ for $N_f=3$ (circle) and 
$N_f=5$ & & & & $N_f=3,\ 5$. \\
(triangle).& & & & \\
\end{tabular}

\vspace{0.5cm}

%\newpage
From Fig. 2,
one sees clearly that at small $n$
$\gamma^{h(1)}_n$ is significantly larger than $\gamma^{f(1)}_n$
but approaches very quickly to $\gamma^{f(1)}_n$, keeping the condition
$\gamma^{h(1)}_n > \gamma^{f(1)}_n$.
This feature is the same as the one-loop anomalous dimensions
$\gamma^{h,f(0)}_n$ which reads
\begin{eqnarray}
\gamma^{f(0)}_n &=& 2C_F \left( 1 - { 2 \over (n+1)(n+2)}
+ 4\sum_{j=2}^{n+1}{1 \over j}\right),
\label{eq4.1}\\
\gamma^{h(0)}_n &=& 2C_F \left( 1 
+ 4\sum_{j=2}^{n+1}{1 \over j}\right),
\label{eq4.2}
\end{eqnarray}
and hence $\gamma^{h(0)}_n > \gamma^{f(0)}_n$ for all $n$.
Actually this tendency is even stronger for the two-loop case.
We note that for $n=0$ the anomalous dimension for $f_1$ is zero
in all orders because of the Ward identity for the vector
current.  On the other hand, $h_1$ projects onto the
tensor operator $\bar{\psi}\sigma^{\mu\nu}\psi$ for $n=0$,
for which there is no conservation law.  
Hence $\gamma^{h(0,1)}_0 \neq 0$.  
From the difference between $\gamma^{f(1)}_n$ and $\gamma^{h(1)}_n$
at small $n$,
we expect that the NLO effect
leads to larger difference
in the $Q^2$-evolution in the small-$x$ region.
We note that the rightmost singularity of $\gamma_n^{h(0)}$
is located at $n=-2$ while that of 
$\gamma_n^{h(1)}$ is at $n=-1$.  Therefore, 
for the case of the transversity
distribution, the NLO evolution renders the DGLAP asymptotics
at $x\to 0$ compatible with the Regge asymptotics\,\cite{KMSS}.

Since the relevant quantities for the $Q^2$-evolution 
of the moments are
$\gamma^{(1)}_n/2\beta_1-\gamma^{(0)}_n/2\beta_0$ (see (\ref{eq2.12})), 
we plotted it in
Fig. 3 for $N_f=3$ and $5$ cases. 
From this figure, one expects that at small $n$ 
the NLO effect in the $Q^2$ evolution is quite different
between ${\cal M}_n[h_1(Q^2)]$ and ${\cal M}_n[f_1(Q^2)]$.
Since $\gamma^{f(0,1)}_n \rightarrow 0$ as $n\to 0$, 
$\gamma^{f(1)}_n/2\beta_1-\gamma^{f(0)}_n/2\beta_0$ abruptly drops to
zero as $n\to 0$.  But this is not the case for $h_1$:  
It shows a characteristic behavior as shown in Fig. 3.

As an example of the $Q^2$-evolution,
we show in Fig. 4 (a) and (b) the $Q^2$-evolution 
of the tensor charge and 
the first moments, respectively, with the parameters $N_f=3$
and $\Lambda=0.232$ GeV in (\ref{eq2.12b}).
(Here we are interested in the NLO effect
in the anomalous dimension and the $\beta$-function, and thus
we adopted the same value for the $\Lambda$-parameter
in the LO and NLO evolution.)

\begin{figure}[h]
%\begin{center}
\epsfile{file=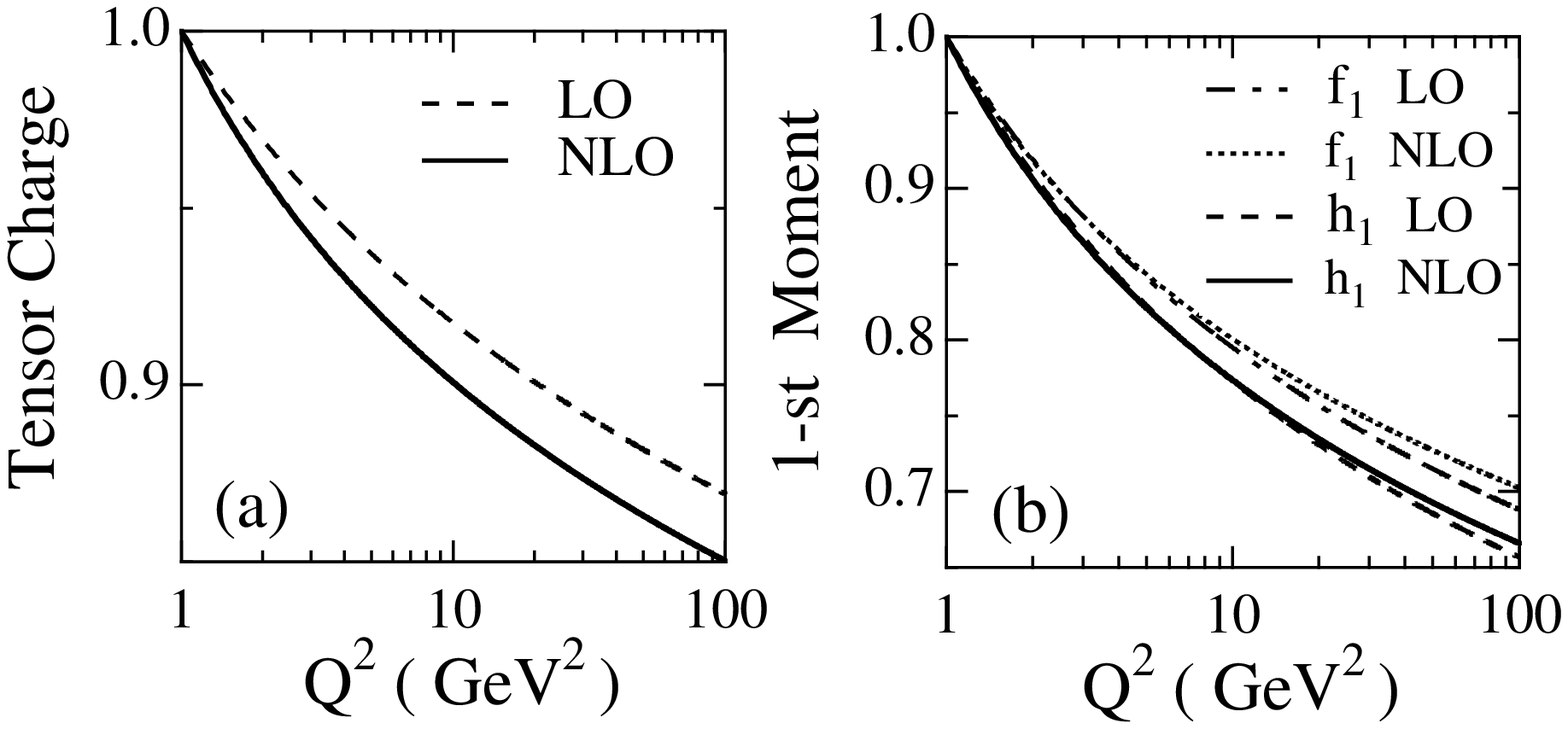,scale=0.8}
%\end{center}
%\caption[]{}
%\protect{\cite{}}}
%\label{figure:fignloC}
\end{figure}

\noindent
Fig. 4: (a) The LO and NLO 
$Q^2$-evolution of the tensor charge
normalized at $Q^2=1$ GeV$^2$. 
(b) The LO and NLO $Q^2$-evolution of the first moment of
$h_1(x,Q^2)$ and $f_1(x,Q^2)$ normalized at $Q^2=1$ GeV$^2$.

\vspace{0.5cm}
They are normalized at $Q^2=1$ GeV$^2$.  
At $n=0$, only diagrams (e), (f), (i), (j), (k) in Fig. 1
survive.  They give
the anomalous dimension for the tensor charge as
\begin{eqnarray}
\gamma^{h(1)}_0 &=& -19C_F^2 +{257\over 9}C_F C_G - 
{52\over 9}C_F T_R\nonumber\\
&=& {724\over 9} -{104\over 27}N_f.
\label{eq4.3}
\end{eqnarray}
From Fig. 4(a),
we can compare the LO and the NLO
$Q^2$-evolution of the tensor charge of the nucleon.
One sees that the NLO effect is sizable 
as is expected from Fig. 3.
In Fig. 4(b), we plotted the LO and NLO $Q^2$-evolution both
for ${\cal M}_1[f_1(Q^2)]$ (nonsinglet) and ${\cal M}_1[h_1(Q^2)]$.
Although the NLO effect in the anomalous dimension
(the second factor in 
the right hand side of (\ref{eq2.12})) makes this ratio 
smaller,
the NLO effect in the coupling constant (the first factor
in the right hand side of (\ref{eq2.12}))
completely cancels this
effect.  For $f_1$, the latter effect is actually 
larger than the former effect.
As is shown in these examples, the NLO effect on each moment
is not large.  We will see, however, that it is significant
for the distribution function itself in the next section.

\section{$Q^2$-evolution of $h_1$}
\setcounter{equation}{0}
\renewcommand{\theequation}{\arabic{section}.\arabic{equation}}

To see a generic feature of the 
actual NLO $Q^2$ evolution of $h_1$, we have applied the result
of the two-loop anomalous dimension
to a reference distribution function of $h_1$.
In the nonrelativistic kinematics, $h_1=g_1$, although the motion
of quarks in the nucleon is not expected to be nonrelativistic.
A bag model calculation suggests $h_1$ is not very different from $g_1$
at a low energy hadronic scale.
We therefore assume that $h_1(x,Q^2)$ is identical to 
a $g_1(x,Q^2)$ distribution of
Gl\"{u}ck, Reya, Stratmann and Vogelsang (GRSV) \cite{GRSV} 
(standard scenario) at a low energy scale.

We first recall the general formalism of the NLO $Q^2$-evolution.
Corresponding to the splitting of a quark into an anti-quark,
there are terms proportional to
$(-1)^{n+1}$ such as those caused from 
$S_k'$ and $\widetilde{S}$
in the anomalous dimension (\ref{eq4.0}).
One can thus write for $R_n(Q^2,\mu^2)$ in (\ref{eq2.12}) as
\beq
R_n(Q^2,\mu^2) = R_n^{qq}(Q^2,\mu^2) + 
(-1)^{n+1}R_n^{q\bar{q}}(Q^2,\mu^2).
\label{eq5.0}
\eeq
We solve the evolution equation (\ref{eq2.12a}) with 
(\ref{eq2.3}) and (\ref{eq5.0}) for $n=$ even and odd cases 
by analitically continuing them to all $n$.
This gives the $Q^2$-evolution for the two combinations
$h_1(x,Q^2) \pm \bar{h}_1(x,Q^2)$.   Numerical solution 
was obtained following the method of \cite{GLY}.
We have checked that the method
works sufficiantly accurately in the $x>0.05$ region
for the NLO evolution.

%\vspace{-0.5cm}
\begin{figure}[h]
%\begin{center}
\epsfile{file=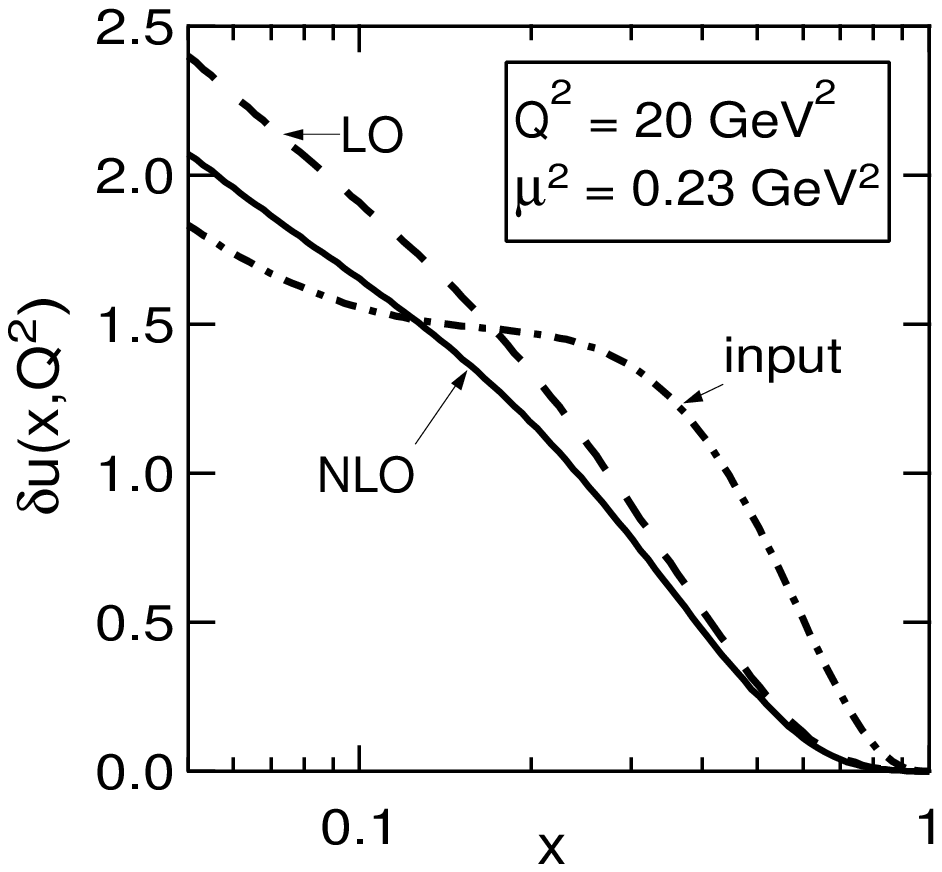,scale=0.7}
%\end{center}
%\caption[]{}
%\protect{\cite{}}}
%\label{figure:fignloC}
\end{figure}

\noindent
Fig. 5: The LO and NLO $Q^2$-evolution for the $u$-quark transversity
distribution $\delta u$, starting from the same input function.

\newpage
\begin{figure}[h]
%\begin{center}
\epsfile{file=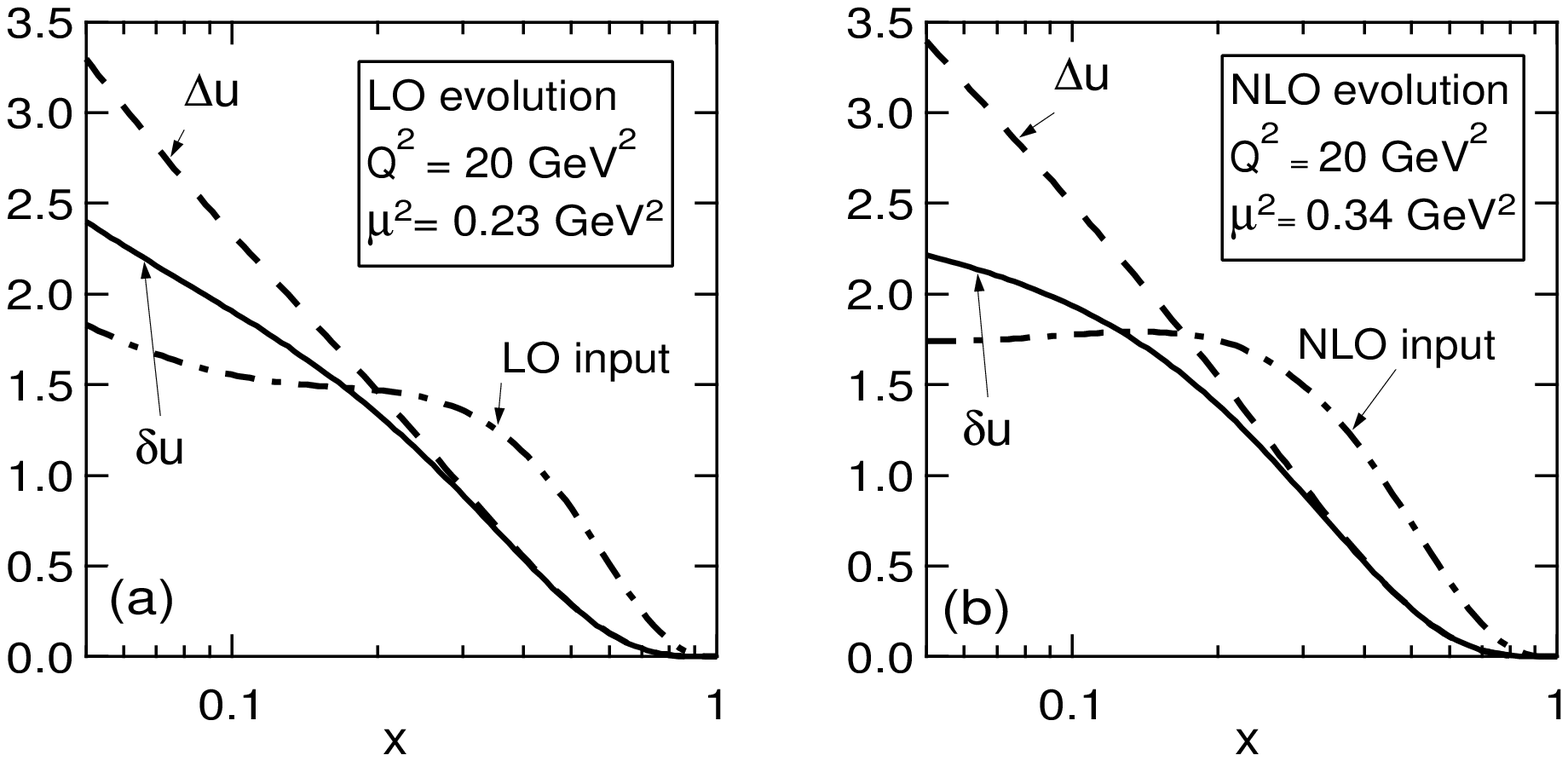,scale=0.7}
%\end{center}
%\caption[]{}
%\protect{\cite{}}}
%\label{figure:fignloC}
\end{figure}
\noindent
Fig. 6: (a) Comparison of the LO $Q^2$-evolution of $\delta u(x,Q^2)$
and $\Delta u(x,Q^2)$.
(b) Comparison of the NLO $Q^2$-evolution of $\delta u(x,Q^2)$
and $\Delta u(x,Q^2)$.

\begin{figure}[h]
%\begin{center}
\epsfile{file=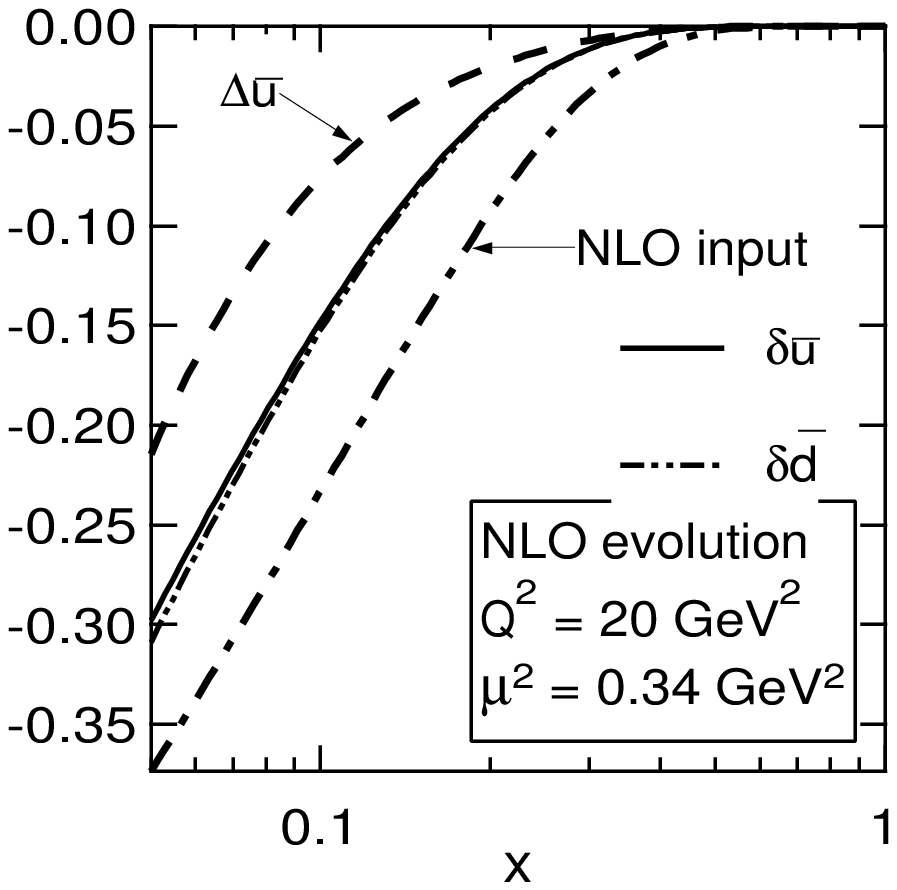,scale=0.7}
%\end{center}
%\caption[]{}
%\protect{\cite{}}}
%\label{figure:fignloC}
\end{figure}
\noindent
Fig. 7.
Comparison of the NLO $Q^2$-evolution of $\delta \bar{u}(x,Q^2)$, 
$\delta \bar{d}(x,Q^2)$ and $\Delta \bar{u}(x,Q^2)$.

\vspace{0.5cm}

To see the magnitude of the NLO effect,
we first applied the LO and NLO evolution of $h_1$
to the same input function.  As an input function
we have chosen here the GRSV LO $g_1$ distribution function
at $\mu^2=0.23$ GeV$^2$, which is the starting distribution
of \cite{GRSV}.
Figure 5 shows the result for the u-quark
transversity distribution $\delta u(x,Q^2)$ evolved to
$Q^2=20$ GeV$^2$.  The NLO effect is significant especially in the 
small-$x$ region.  The same tendency was observed for the other 
transversity distributions.

To compare the $Q^2$-evolution of the transversity distribution and
the $g_1$-distribution,
we show in Fig. 6 (a) and (b), respectively, 
the LO and NLO $Q^2$-evolution of $\delta u$ and $\Delta u$
($g_1$ for $u$-quark), starting from
the GRSV LO and NLO input function 
for $\Delta u$ given at $Q^2=0.23$ and $0.34$ GeV$^2$.
Although the LO evolution leads to a significant difference
between $\delta u$ and $\Delta u$ at $Q^2= 20$ GeV$^2$,
\footnote{This difference was also pointed out in \cite{BCD}.}
this feature is even more conspicuous in the NLO
evolution, in particular, in the small-$x$ region.
This is expected from the behavior of the anomalous dimension.
(Although the evolution of $\Delta u$ shown in Fig. 6 is receiving some 
effect from the mixing with the gluon distribution, flavor-nonsinglet
and valence distributions also show the same tendency.)

In Fig. 7, we compare the NLO $Q^2$-evolution of $\delta \bar{u}$,
$\delta \bar{d}$ and
$\Delta \bar{u}$, starting from the same input distribution function
(GRSV NLO input function for the sea quark distribution for $g_1$).
The difference between $\delta \bar{u}$ and 
$\Delta \bar{u}$ is again significant.
Although the input sea quark distribution is
assumed to be flavor symmetric ($\delta\bar{u} = \delta\bar{d}$
at $Q^2=0.34$ GeV$^2$),  
the NLO evolution leads to a violation of this symmetry
because of the
presence of $R_n^{q\bar{q}}(Q^2,\mu^2)$ in (\ref{eq2.12}).
However, this effect is very small as can be seen from Fig. 7.
\footnote{This point is also discussed in \cite{MSSV}.  
At the time of the workshop, we reported this effect
is sizable for $h_1$ compared with $g_1$ case.  However,
our numerical solution at that time turned out to be unreliable
in the small $x$ region ($x<0.05$). 
}

\section{Summary and outlook}
\setcounter{equation}{0}
\renewcommand{\theequation}{\arabic{section}.\arabic{equation}}

In this work, we have carried out a Feynmann gauge calculation of
the two-loop anomalous dimension for the transversity 
distribution $h_1(x,Q^2)$ in the 
MS scheme of the dimensional regularization.   
This completes the calculation of the anomalous dimensions
for all the twist-2 distributions of the nucleon in the NLO level.
We found $\gamma^{h(1)}_n$ is significantly larger 
than the $\gamma^{f,g(1)}_n$ (nonsinglet anomalous dimension for 
$f_1$ and $g_1$)
at small $n$, but approaches very quickly to $\gamma^{f,g(1)}_n$
at large $n$, keeping the condition
$\gamma^{h(1)}_n >\gamma^{f,g(1)}_n$.  
The NLO effect in the anomalous dimension
leads to a drastic difference in the
$Q^2$-evolution between $h_1$ and $g_1$.
The NLO parton distribution function 
has to be combined 
with the NLO short distance cross section 
in the same factorization scheme to give
a prediction for a physical quantity.
For the transversely polarized Drell-Yan process, the latter has been 
calculated in \cite{Vog,CKM}.
Combining these results,  
the NLO calculation of the double spin asymmetry for the Drell-Yan
process has been presented
in a recent work \cite{MSSV}.
We hope the peculiar feature of the $h_1$ distribution discussed
in these works will be measured in the future collider experiments.

\vspace{0.3cm}

\cl{\bf Acknowledgement}

This work is supported in part by RIKEN.  We thank M. Stratmann and
W. Vogelsang for providing us with the Fortran code of their
$g_1$ distribution. 

%\vspace{0.5cm}

%\newpage

\end{document}